\documentclass[aps,prb,10pt,twocolumn,amsmath,showpacs,superscriptaddress,reprint]{revtex4-2} 
\usepackage[hmargin=2cm,vmargin=2.5cm]{geometry}

\usepackage[version=4]{mhchem} 
\usepackage{verbatim}

\usepackage[colorlinks,plainpages=false,linkcolor=blue,urlcolor=blue,citecolor=blue,pdfpagemode=UseNone,pdfstartview=FitBH]{hyperref} 
\usepackage[Charter,greekuppercase=italicized]{mathdesign} 
\usepackage{graphicx}    
\graphicspath{{./}{figs/}}
\DeclareGraphicsExtensions{.eps,.png,.pdf,.jpg}

\newcommand{\nbs}{\ce{NbSe2}}
\newcommand{\snbs}{Sn-intercalated \ce{NbSe2}}
\newcommand{\vxv}{$\sqrt{13}\times\sqrt{13}$}

\begin{document}

\title{Emergence of a correlated insulating state in bulk 1T-{\nbs} via metal intercalation}
\author{M.~Tomlinson}
\author{AKM A.~Rahman}
\author{S.~Devi}
\author{R.~Tuchikawa}
\altaffiliation[Present address: ]{Truventic LLC, Winter Park, Florida, 32789, USA}
\author{M.~Ishigami}
\author{D.~Le}
\affiliation{Department of Physics, University of Central Florida, Orlando, Florida 32816, USA}
\author{Md Z.~Mohayman}
\affiliation{Department of Materials Science and Engineering, University of Central Florida, Orlando, Florida 32816, USA}
\author{A.~Kushima}
\affiliation{Department of Materials Science and Engineering, University of Central Florida, Orlando, Florida 32816, USA}
\affiliation{Advanced Materials Processing and Analysis Center, University of Central Florida, Orlando, Florida 32816, USA}
\author{Y.~Nakajima}
\email[Corresponding author: ]{Yasuyuki.Nakajima@ucf.edu}
\affiliation{Department of Physics, University of Central Florida, Orlando, Florida 32816, USA}

%\date{\today}

\date{\today}

\begin{abstract}
  The 1T polymorph of \nbs, long confined to the monolayer limit, has remained inaccessible in bulk. Here, we report the realization of bulk 1T-\nbs\ via electrochemical Sn intercalation. Transmission electron microscopy directly reveals the formation of the 1T structure induced by Sn intercalation. The intercalated samples exhibit insulating transport behavior, in stark contrast to metallic 2H-\nbs. Density functional theory calculations, however, predict a metallic band structure, highlighting the crucial role of emergent electronic correlations in the observed insulating state. Raman spectroscopy further reveals vibrational modes associated with Sn intercalation and possible charge density wave order. Our results establish electrochemical intercalation as an effective route to stabilize otherwise inaccessible bulk polytypes, positioning bulk 1T-\nbs\ as a new platform for investigating correlated electronic states.
\end{abstract}

% \keywords{}

\maketitle

\section{Introduction}
Transition metal dichalcogenides (TMDs) host a wide range of emergent electronic phases enabled by their structural polymorphism. Distinct coordinations—trigonal prismatic H phase and octahedral T phase—stabilize markedly different electronic states, ranging from superconductivity to correlation-driven insulating behavior. Among these materials, \nbs\ is a prototypical system in which intertwined orders emerge. The bulk 2H phase [FIG.1 (a)] undergoes a charge density wave (CDW) transition at 33 K \cite{monct77,kiss07}, followed by superconductivity below 7.2 K \cite{harpe75}. The monolayer 1H phase hosts Ising superconductivity, in which electron spins are locked in the out-of-plane direction due to spin-orbit coupling, protected against in-plane magnetic fields \cite{xi16}. Additional stacking variants, such as 3R and 4H phases, further enrich this phase space with noncentrosymmetric superconductivity \cite{li26} and possible bulk Ising superconductivity \cite{patra25,marti26}.

In contrast, the octahedrally coordinated 1T phase of \nbs [FIG.1 (b)]—believed to harbor strong correlation effects—has remained inaccessible in bulk form and has only been observed in monolayer islands \cite{nakat16,nakat21,liu21d,liu21c,zhang24b}. In the molecular-beam-eptaxy-grown monolayer \nbs, scanning tunneling microscopy (STM) reveals a $\sqrt{13}\times\sqrt{13}$ star-of-David CDW accompanied by a Mott gap, suggesting a correlated insulating state \cite{nakat16,nakat21,liu21d,liu21c}. Another STM study of magnetic molecules deposited on the monolayer 1T-\nbs\ elucidate the spinon-Kondo effect, pointing toward the realization of a quantum spin liquid state \cite{zhang24b}. Thus, realizing the 1T phase in bulk \nbs\ would provide a new platform for exploring correlation-driven phenomena beyond the two-dimensional limit. A promising route to this regime is electrochemical intercalation, which can drive structural transformations, such as the 2H-to-1T transition in \ce{MoS2} \cite{shi24}.

Here we report that electrochemical Sn intercalation stabilizes the 1T phase of \nbs\ in bulk form. X-ray diffraction (XRD) reveals a systematic expansion of the interlayer spacing, consistent with Sn intercalation in \nbs. Transmission electron microscopy (TEM) directly confirms the emergence of the 1T structure in the \snbs. Transport measurements show insulating behavior at low temperatures without a clear phase transition between 2 and 300 K, reminiscent of Mott insulating TMDs such as 1T-\ce{TaS2}. In contrast, density functional theory (DFT) predicts a metallic band structure in the bulk 1T-\nbs, indicating that electronic correlations are essential to understanding the observed insulating state. Raman spectroscopy further identifies additional vibrational modes, associated with intercalant-induced distortions and/or a possible CDW formation. These results establish electrochemical Sn intercalation as an effective route to stabilize otherwise inaccessible bulk polytypes in TMDs. The realization of bulk 1T-\nbs\ opens a pathway to investigate correlated insulating states and related quantum phases in a tunable, three-dimensional platform.

%Figure for XRD
\begin{figure*}[t]
\includegraphics[width=0.95\linewidth]{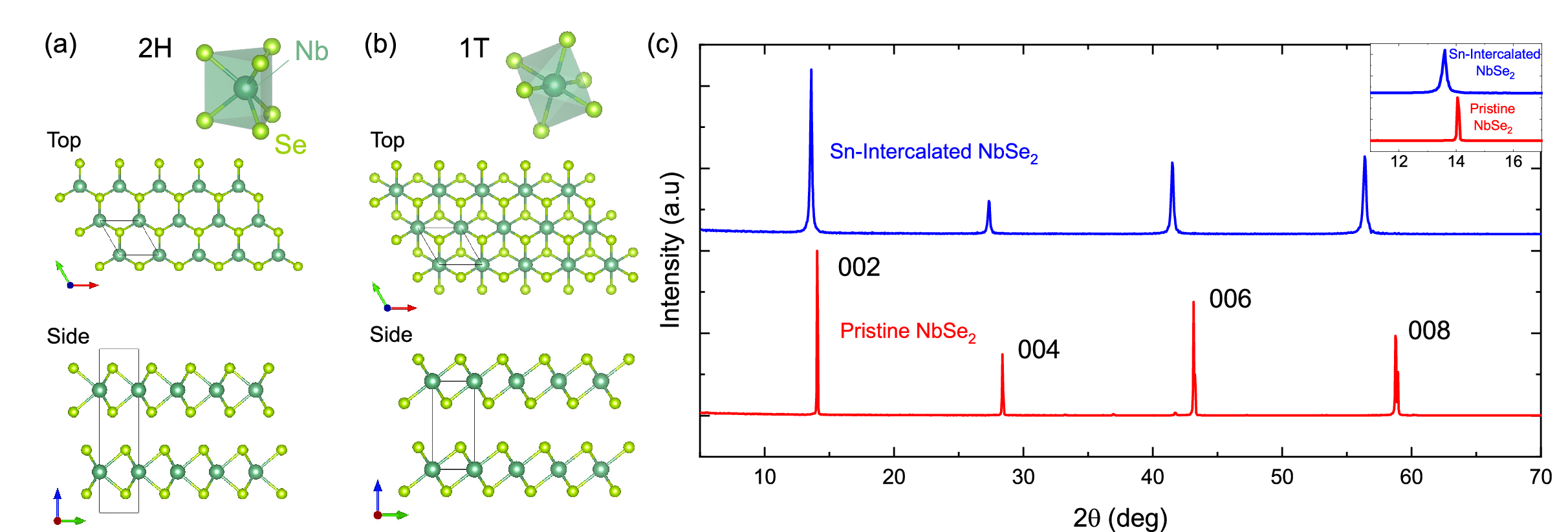}
\caption{\label{fig:xrd} Coordinations and crystal structures of (a) 2H-\nbs\ and (b) 1T-\nbs. (c) XRD patterns of Sn-intercalated and pristine \nbs\ using a Cu $K_\alpha$ radiation. Inset: XRD patterns for Sn-intercalated and pristine \nbs\ centered around 14$^{\circ}$.}
\end{figure*}

\section{Methods}
%\subsubsection{Sample Preparation}
Polycrystalline \nbs\ was synthesized via a solid-state reaction. Elemental Nb and Se in a stoichiometric ratio were heated in a sealed quartz ampule at 850~$^{\circ}$C for 5 days. High-quality single crystals were subsequently grown by chemical vapor transport using I$_{2}$ as the transport agent. The polycrystalline material sealed in a quartz ampule with 3~mg/cm$^3$ of I$_{2}$ was placed in a two-zone furnace at 850~$^{\circ}$C and 800~$^{\circ}$C for 21 days.

For electrochemical intercalation, the as-grown \nbs\ crystals were wrapped in high-purity Sn wire serving as the working electrode, with a second Sn wire as the counter electrode. A leak-free Ag/AgCl electrode was used as the reference. Intercalation was performed at a constant potential of $-0.7$~V at room temperature for 2~h, using 0.1\% HCl as the electrolyte. The Sn content in the intercalated samples was determined by X-ray fluorescence (XRF) spectroscopy.

Electrical resistivity was measured in the ab-plane with a standard four-wire configuration using an AC resistance bridge. An excitation current of 316~$\mu$A was applied. Raman spectra were collected using a 532~nm laser. 

% TEM
High-resolution transmission electron microscopy (HRTEM) analysis was performed using Tecnai F30 (FEI) operated at 300 kV acceleration voltage. TEM sample was prepared by attaching a piece of synthesized NbSe$_{2}$ sample on a Cu TEM grid using conductive epoxy.  Peeling off the sample after curing the epoxy leaves thin flakes of NbSe$_{2}$ on the grid for the observation.

% DFT
Spin-polarized DFT calculations are performed using the Vienna Ab initio Simulation Package (VASP 5.4.4)\cite{kress93,kress96}, employing the projector-augmented wave (PAW) pseudopotential method and a plane wave basis set \cite{bloch94,kress99}. We use the generalized-gradient approximation (GGA) in the form of Perdew-Berke-Enzerhoff (PBE) functionals \cite{perde96,perde97} with a Hubbard $U$ of 3 eV for Nb 4$d$ electrons for describing the exchange correlation \cite{liu21c,liu24c}. DFT-D3 correction was used for accounting for the van der Waals (vdW) interactions \cite{grimm10}. We set a cutoff energy of 500 eV for plane-wave expansion. All electronic iterations are converged with 0.01 meV threshold. The internal coordinate of atoms and lattice constant of the bulk structures are optimized so that forces acting on each atom are less than 5 meV/\AA\ and the stress is minimized to less than $0.2k_{\mathrm{B}}$. We model bulk structures \nbs with lateral dimension of \vxv\ which include two layers of \nbs (26 Nb and 52 Se atoms). We chose this supercell to accommodate the potential \vxv\ CDW reconstruction of 1T-\nbs.  For modeling Sn intercalation, we used one Sn between the \nbs\ layers, i.e., $\sim$7.7\% Sn intercalation. We use Gaussian smearing method with $\sigma$ = 0.1 eV and sample Brillouin Zone with a 3$\times$3$\times$3 $\Gamma$-centered grid.

\section{Results and Discussion} 
% Results for XRD
We first confirm that the crystal structure of pristine \nbs\ is consistent with the 2H polytype (space group $P6_{3}/mmc$), as shown in Fig.~\ref{fig:xrd}(c). The refined lattice constant is $c = 12.5668(7)$~\AA, in good agreement with previous reports \cite{meyer75,naik19}. Notably, while the XRD pattern of the Sn-intercalated sample closely resembles that of pristine \nbs, all peaks shift systematically toward lower $2\theta$, as highlighted in the inset of Fig.~\ref{fig:xrd} (c). This shift indicates a significant expansion of the interlayer spacing, suggesting the successful intercalation of Sn in the vdW gap.

XRF measurements determine the intercalated Sn concentration to be approximately 6\%. Assuming that the intercalated sample retains the 2H structure, we obtain an effective lattice constant of $c = 13.0340(8)$~\AA, corresponding to an expansion rate of $\sim 0.08$~\AA/\%. This value is nearly an order of magnitude larger than the previously reported expansion rate of $\sim 0.01$~\AA/\% for Sn intercalation in polycrystalline 2H-\nbs\ \cite{naik19}. The anomalously large lattice expansion observed in our sample strongly suggests that Sn is not simply intercalated within the 2H phase. Instead, it points to an intercalation-driven structural transformation into a distinct phase.

\begin{figure}[t]
f\includegraphics[width=0.95\linewidth]{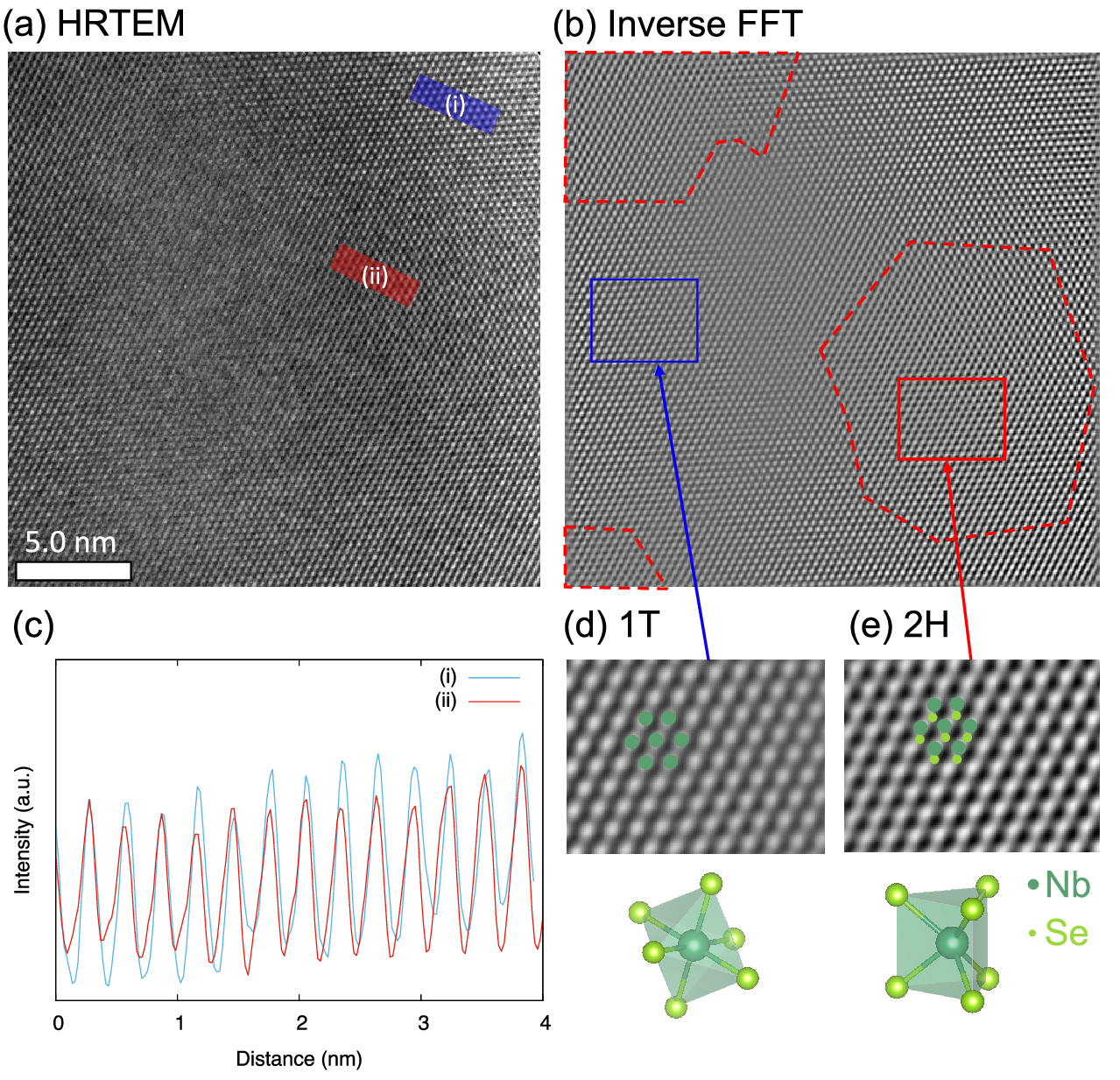}
\caption{\label{fig:TEM}(a) High-resolution TEM and (b) its inverse FFT images for pristine and Sn-intercalated \nbs. (c) Line profiles from regions (i) and (ii), highlighted in (a), corresponding to the 1T and 2H phases, respectively. Zoomed-in images of (d) the 1T phase and (e) the 2H phase, corresponding to the blue- and red-square regions in (b), respectively.}
\end{figure}

Indeed, TEM directly confirms the structural transformation to the 1T phase driven by Sn intercalation. High-resolution TEM and its inverse fast Fourier transform (FFT) images (Fig.~\ref{fig:TEM}) reveal the characteristic trigonal arrangement of Nb atoms in the 1T structure, with minor regions of the 2H phase exhibiting the hexagonal atomic arrangement, surrounded by the red dashed lines in the figure. In addition, lattice parameter a of the in-plane hexagonal lattice of the \nbs, obtained from the line profiles (i) and (ii) in FIG.\ref{fig:TEM}(a), were measured to be 3.42 \AA\ and 3.39~\AA\ for 1T and 2H regions, respectively. These observations provide direct evidence that electrochemical Sn intercalation drives a structural transition from 2H to 1T in bulk \nbs.

\begin{figure}[t]
\includegraphics[width=0.95\linewidth]{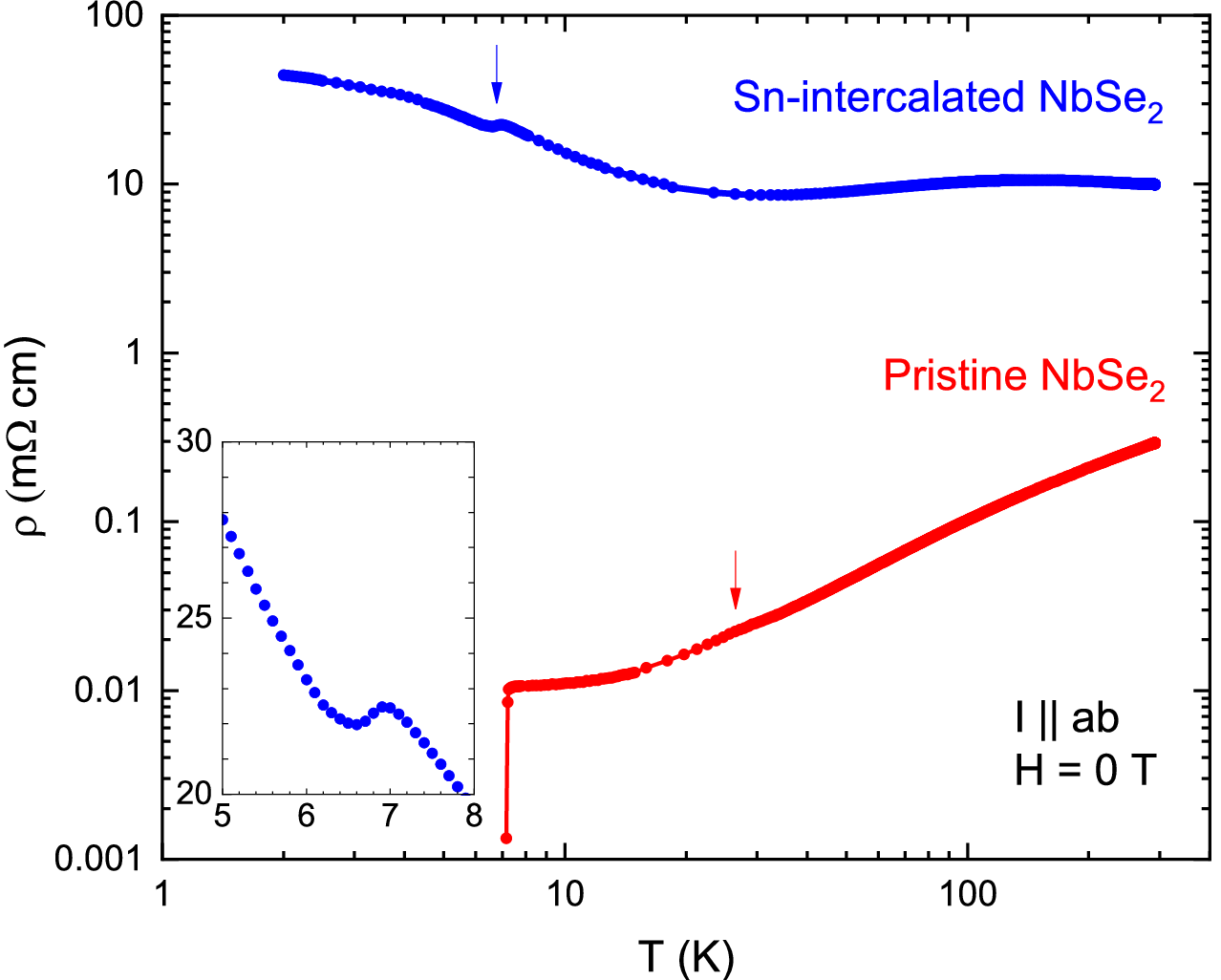}
\caption{\label{fig:res}Temperature dependence of the resistivity in the ab-plane for pristine and Sn-intercalated \nbs. The inset displays low temperature resistivity of \snbs\, highlighting a transition at 7 K.}
\end{figure}

% Results for Resistivity
Electrochemical Sn intercalation into the vdW layers of \nbs\ has a pronounced impact on its charge transport. Pristine \nbs\ in the 2H structure exhibits metallic behavior upon cooling, as shown in FIG.~\ref{fig:res}. At $T=$ 32 K, indicated by the red arrow, the resistivity displays an anomaly associated with the CDW transition, followed by a superconducting transition at $T_c = 7.2$ K. The resistivity values at 290 K and $T_c$ are $\sim$290~$\mu\Omega$\,cm and $\sim$10~$\mu\Omega$\,cm, respectively, yielding a residual resistivity ratio $\rho(290\,\mathrm{K})/\rho(T_c) \approx 28$.

%Figure for Resistivity

In contrast, the \snbs\ exhibits dramatically enhanced resistivity, reaching $\sim$10~m$\Omega$\,cm at 290 K, approximately 34 times larger than that of the pristine sample. Upon cooling, the resistivity shows insulating behavior below $\sim$30 K, after showing a broad hump centered around 150 K. A small peak is observed near 7 K, indicated by the blue arrow, which we attribute to a superconducting transition arising from a minor fraction of remaining 2H-\nbs\ regions, observed in the TEM image. This minority phase is likely below the detection limit of XRD, consistent with the absence of corresponding peaks in FIG.~\ref{fig:xrd}.

The observed insulating transport cannot be attributed to disorder-induced localization. In particular, Anderson localization arising from impurity scattering would require sufficiently strong disorder within the conducting TMD planes \cite{disa76}. However, the Sn intercalants reside in the vdW gaps between layers and therefore couple only weakly to in-plane electronic motion. Consistent with this picture, in Sn-intercalated 2H-\nbs\ a comparable intercalation level (4\%) increases residual resistivity but preserves metallic transport without any indication of localization \cite{naik19}. The stark contrast with the insulating behavior observed here thus rules out disorder as the primary origin and instead points to other effects.

Rather, the insulating behavior observed in \snbs\ is reminiscent of the Mott insulating state in 1T-\ce{TaS2}. In 1T-\ce{TaS2}, the resistivity increases upon cooling from room temperature and exhibits a pronounced jump at the commensurate CDW transition, followed by insulating behavior at low temperatures \cite{wilso75a}. This behavior is attributed to electron localization associated with the formation of a star-of-David CDW superstructure with $\sqrt{13}\times\sqrt{13}$ periodicity \cite{law17}. In contrast, although \snbs\ shows a similar overall insulating behavior, no clear anomaly associated with such a transition is observed within the measured temperature range up to 300 K, suggesting that the CDW transition may occur at higher temperatures. Our observations suggests the emergence of a correlated insulating state realized in the 1T-phase of bulk \snbs.

\begin{figure}[t]
\includegraphics[width=0.95\linewidth]{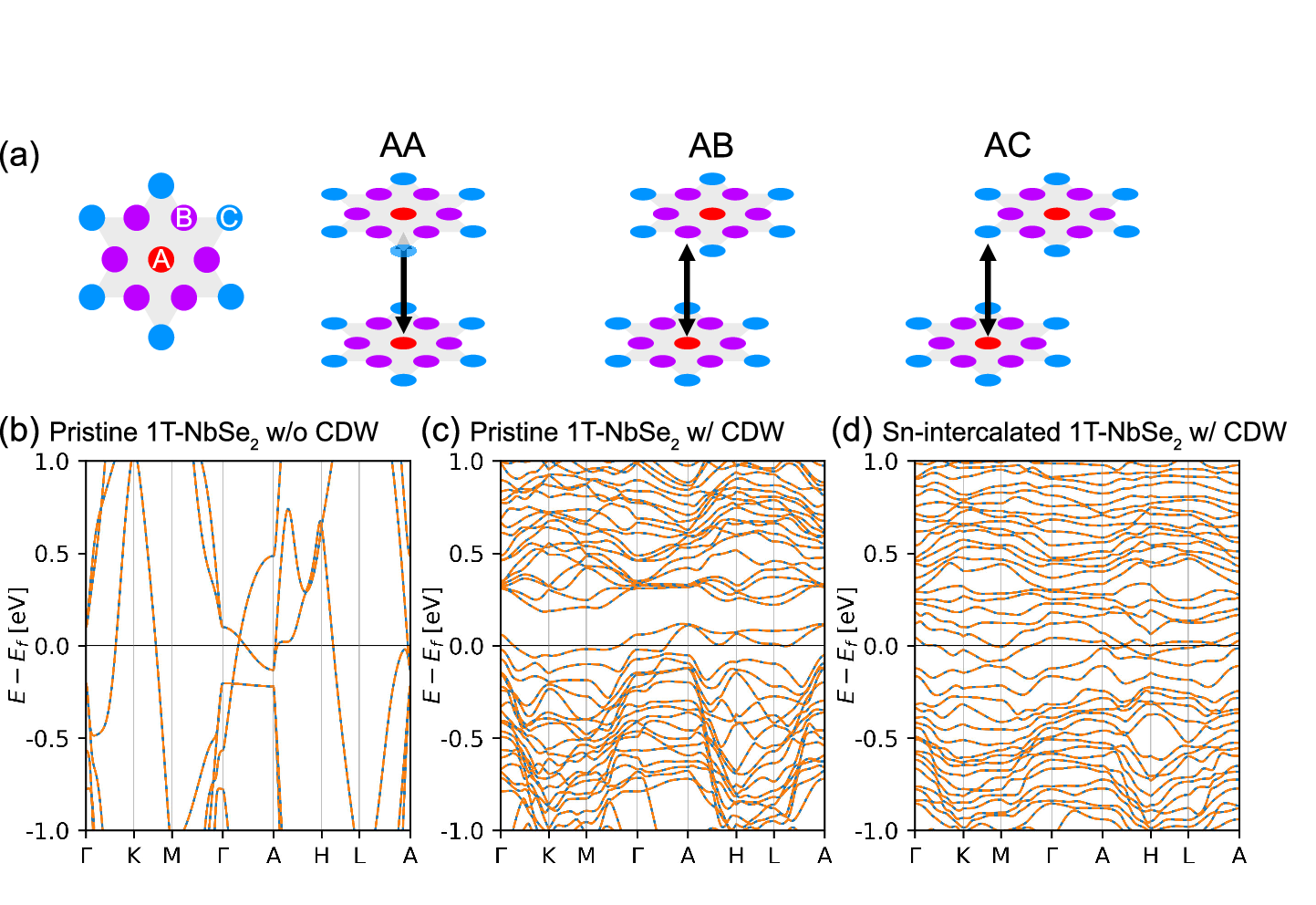}
\caption{\label{fig:bands}(a) Schematic of the $c$-axis stacking order of the star-of-David units in the CDW phase of 1T-\nbs. Band structures of (b) pristine 1T-\nbs\ without CDW (nonreconstructed), (c) pristine 1T-\nbs\ with the \vxv\ CDW (reconstructed), and (d) Sn-intercalated 1T-\nbs\ with the \vxv\ CDW (reconstructed). Blue-solid and orange-dashed lines represent spin-up and spin-down bands, respectively.}
\end{figure}

% Results for DFT
Despite the observed insulating behavior, DFT calculations predict a metallic band structure for Sn-intercalated 1T-\nbs. As shown in Fig.~\ref{fig:bands}(a), bulk pristine 1T-\nbs\ without the \vxv\ CDW exhibits a metallic band structure. It has been suggested that the stacking of the star-of-David CDW plays a crucial role in determining the ground state of 1T-TMDs \cite{zhang23c}. To investigate this effect, we construct several stacking configurations based on the relative alignment of the star-of-David pattern. Specifically, we consider three geometries: (i) AA stacking, where the centers A of the star-of-David units are vertically aligned; (ii) AB stacking, where an inner vertex B of one star-of-David lies above the center A of a star-of-David in the adjacent layer; and (iii) AC stacking, where an outer vertex C lies above the center A in the neighboring layer, as illustrated in Fig.~\ref{fig:bands}(a). Among these, AC stacking has the lowest total energy for bulk pristine 1T-\nbs. As shown in Fig.~\ref{fig:bands}(b), the reconstructed band structure with the AC stacking CDW order is either metallic or semiconducting, depending on the position of the Fermi level. This suggests that a shift in the Fermi energy induced by Sn intercalation could, in principle, render the system semiconducting.

However, Sn intercalation does not open a gap, and the system remains metallic, according to our calculations. To calculate the band structure of \snbs, we start with AC stacking, which has the lowest total energy. We model Sn intercalation by placing one Sn atom above the center A of a star-of-David unit in one \nbs\ layer, while the second Sn atom is positioned above either the center A, an inner vertex B, or an outer vertex C of a star-of-David unit in the adjacent layer. Among these three intercalation configurations, the lowest-energy structure corresponds to the case in which the second Sn atom is located above the inner vertex B of the star of David. Notably, as shown in FIG.\ref{fig:bands}(c), the band structure with the Sn-intercalated 1T-\nbs\ with the CDW is metallic. This discrepancy between experiments and calculations indicates that electronic correlations may play a crucial role in stabilizing the observed insulating state. Further theoretical work is required to clarify the nature of these correlation effects.

%Results for Raman
Raman spectroscopy further supports Sn intercalation in \nbs. For pristine \nbs, the measured Raman peaks are consistent with previous reports \cite{perei82}. The irreducible representation at the Brillouin zone center for 2H-\nbs\ (point group $D_{6h}$) is \cite{perei82}:
\begin{eqnarray}
  \begin{split}
  \Gamma = 2A_{2u}+2B_{2g}+B_{1u}+A_{1g}+2E_{1u}\\
    +2E_{2g}+E_{2u}+E_{1g}.
    \end{split}
\end{eqnarray}
Among these, the Raman-active modes are one $A_{1g}$, one $E_{1g}$, and two $E_{2g}$. In our backscattering configuration from the basal plane, $A_{1g}$ and $E_{2g}$ modes are expected to dominate. Reported values are 29~cm$^{-1}$ for $E_{2g}^{2}$, 228~cm$^{-1}$ for $A_{1g}$, and 237~cm$^{-1}$ for $E_{2g}^{1}$. Our observed peak around 232~cm$^{-1}$ corresponds to overlapping $A_{1g}$ and $E_{2g}^{1}$ modes. The $E_{1g}$ mode is not observed due to the measurement geometry, and the broad feature around 190~cm$^{-1}$ arises from a two-phonon process, also consistent with previous reports.
In \snbs, we observe three peaks at 122, 227, and 257~cm$^{-1}$. For 1T-\nbs\ (point group $D_{3d}$), the Brillouin zone center representation is
\begin{eqnarray}
\Gamma = A_{1g}+2A_{2u}+E_{g}+A_{1g}+2E_{u},
\end{eqnarray}
with one $A_{1g}$ and one $E_g$ mode Raman-active. According to our DFT calculations, the peak at 227~cm$^{-1}$ corresponds to the $A_{1g}$ mode, while the $E_g$ mode is not observed due to weak intensity. The additional peaks at 122 and 257~cm$^{-1}$ are absent in the calculated spectrum for pristine 1T-\nbs, suggesting they originate from vibrational modes associated with the intercalated Sn and/or possible CDW order.

%Figure for Raman

\begin{figure}[t]
\includegraphics[width=0.95\linewidth]{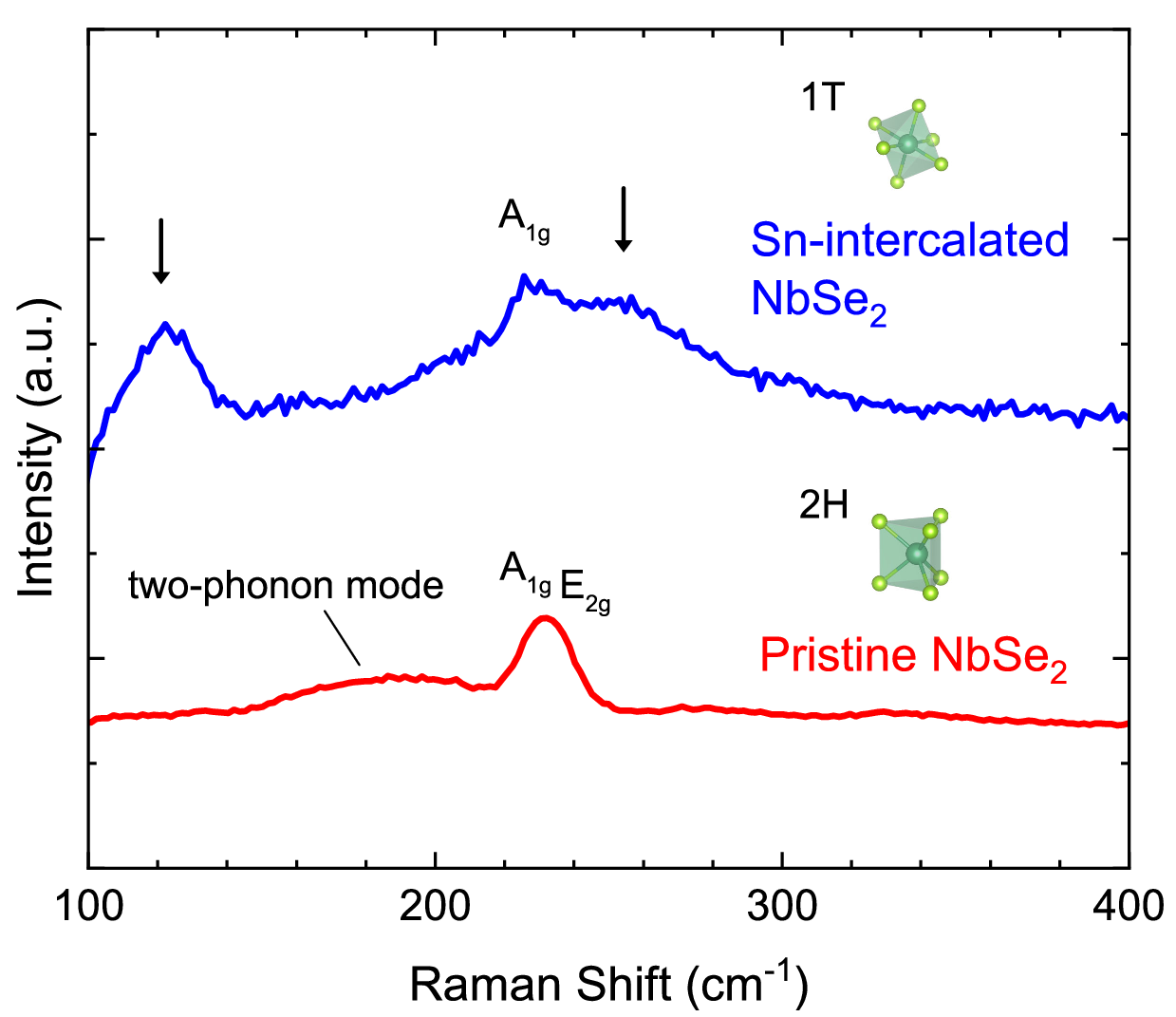}
\caption{\label{fig:raman}Raman spectra for pristine and Sn-intercalated \nbs. Arrows indicate peaks absent from the calculated Raman-active modes for pristine 1T-\nbs.}
\end{figure}

\section{Summary}
We have synthesized the Sn-intercalated \nbs. The TEM measurements clearly reveal the presence of the 1T phase of \nbs\ in the bulk form. The resistivity measurements show insulating behavior at low temperatures, similar to 1T-\ce{TaSe2}. In spite of the insulating behavior, DFT calculations predict that the 1T-\nbs is a metal, suggesting that electron correlations are key ingredients to the observed insulating behavior in \snbs. Raman spectroscopy suggests the additional peaks to the 1T-phase due to the vibrational modes of intercalants and/or a possible CDW phase. Our results suggest that electrochemical Sn intercalation stabilizes the unrealized bulk 1T-\nbs, providing a new platform for correlated physics.

\begin{acknowledgments}
M.T.\ and Y.N.\ were supported by an NSF Career DMR-1944975.  
\end{acknowledgments}

\bibliographystyle{apsrev4-2_YN.bst}

\begin{thebibliography}{28}%
\makeatletter
\providecommand \@ifxundefined [1]{%
 \@ifx{#1\undefined}
}%
\providecommand \@ifnum [1]{%
 \ifnum #1\expandafter \@firstoftwo
 \else \expandafter \@secondoftwo
 \fi
}%
\providecommand \@ifx [1]{%
 \ifx #1\expandafter \@firstoftwo
 \else \expandafter \@secondoftwo
 \fi
}%
\providecommand \natexlab [1]{#1}%
\providecommand \enquote  [1]{``#1''}%
\providecommand \bibnamefont  [1]{#1}%
\providecommand \bibfnamefont [1]{#1}%
\providecommand \citenamefont [1]{#1}%
\providecommand \href@noop [0]{\@secondoftwo}%
\providecommand \href [0]{\begingroup \@sanitize@url \@href}%
\providecommand \@href[1]{\@@startlink{#1}\@@href}%
\providecommand \@@href[1]{\endgroup#1\@@endlink}%
\providecommand \@sanitize@url [0]{\catcode `\\12\catcode `\$12\catcode
  `\&12\catcode `\#12\catcode `\^12\catcode `\_12\catcode `\%12\relax}%
\providecommand \@@startlink[1]{}%
\providecommand \@@endlink[0]{}%
\providecommand \url  [0]{\begingroup\@sanitize@url \@url }%
\providecommand \@url [1]{\endgroup\@href {#1}{\urlprefix }}%
\providecommand \urlprefix  [0]{URL }%
\providecommand \Eprint [0]{\href }%
\providecommand \doibase [0]{http://dx.doi.org/}%
\providecommand \selectlanguage [0]{\@gobble}%
\providecommand \bibinfo  [0]{\@secondoftwo}%
\providecommand \bibfield  [0]{\@secondoftwo}%
\providecommand \translation [1]{[#1]}%
\providecommand \BibitemOpen [0]{}%
\providecommand \bibitemStop [0]{}%
\providecommand \bibitemNoStop [0]{.\EOS\space}%
\providecommand \EOS [0]{\spacefactor3000\relax}%
\providecommand \BibitemShut  [1]{\csname bibitem#1\endcsname}%
\let\auto@bib@innerbib\@empty
%</preamble>
\bibitem [{\citenamefont {Moncton}\ \emph {et~al.}(1977)\citenamefont
  {Moncton}, \citenamefont {Axe},\ and\ \citenamefont {DiSalvo}}]{monct77}%
  \BibitemOpen
  \bibfield  {author} {\bibinfo {author} {\bibfnamefont {D.~E.}\ \bibnamefont
  {Moncton}}, \bibinfo {author} {\bibfnamefont {J.~D.}\ \bibnamefont {Axe}}, \
  and\ \bibinfo {author} {\bibfnamefont {F.~J.}\ \bibnamefont {DiSalvo}},\
  }\bibfield  {title} {\bibinfo {title} {Neutron scattering study of the
  charge-density wave transitions in 2H-$\mathrm{Ta}{\mathrm{Se}}_{2}$ and
  2H-$\mathrm{Nb}{\mathrm{Se}}_{2}$},\ }\href {\doibase
  10.1103/PhysRevB.16.801} {\bibfield  {journal} {\bibinfo  {journal} {Phys.
  Rev. B}\ }\textbf {\bibinfo {volume} {16}},\ \bibinfo {pages} {801} (\bibinfo
  {year} {1977})}\BibitemShut {NoStop}%
\bibitem [{\citenamefont {Kiss}\ \emph {et~al.}(2007)\citenamefont {Kiss},
  \citenamefont {Yokoya}, \citenamefont {Chainani}, \citenamefont {Shin},
  \citenamefont {Hanaguri}, \citenamefont {Nohara},\ and\ \citenamefont
  {Takagi}}]{kiss07}%
  \BibitemOpen
  \bibfield  {author} {\bibinfo {author} {\bibfnamefont {T.}~\bibnamefont
  {Kiss}}, \bibinfo {author} {\bibfnamefont {T.}~\bibnamefont {Yokoya}},
  \bibinfo {author} {\bibfnamefont {A.}~\bibnamefont {Chainani}}, \bibinfo
  {author} {\bibfnamefont {S.}~\bibnamefont {Shin}}, \bibinfo {author}
  {\bibfnamefont {T.}~\bibnamefont {Hanaguri}}, \bibinfo {author}
  {\bibfnamefont {M.}~\bibnamefont {Nohara}}, \ and\ \bibinfo {author}
  {\bibfnamefont {H.}~\bibnamefont {Takagi}},\ }\bibfield  {title} {\bibinfo
  {title} {Charge-order-maximized momentum-dependent superconductivity},\
  }\href {\doibase 10.1038/nphys699} {\bibfield  {journal} {\bibinfo  {journal}
  {Nature Physics}\ }\textbf {\bibinfo {volume} {3}},\ \bibinfo {pages} {720}
  (\bibinfo {year} {2007})}\BibitemShut {NoStop}%
\bibitem [{\citenamefont {Harper}\ \emph {et~al.}(1975)\citenamefont {Harper},
  \citenamefont {Geballe},\ and\ \citenamefont {{Di Salvo}}}]{harpe75}%
  \BibitemOpen
  \bibfield  {author} {\bibinfo {author} {\bibfnamefont {J.}~\bibnamefont
  {Harper}}, \bibinfo {author} {\bibfnamefont {T.}~\bibnamefont {Geballe}}, \
  and\ \bibinfo {author} {\bibfnamefont {F.}~\bibnamefont {{Di Salvo}}},\
  }\bibfield  {title} {\bibinfo {title} {Heat capacity of 2H-NbSe$_{2}$ at the
  charge density wave transition},\ }\href {\doibase
  https://doi.org/10.1016/0375-9601(75)90592-7} {\bibfield  {journal} {\bibinfo
   {journal} {Physics Letters A}\ }\textbf {\bibinfo {volume} {54}},\ \bibinfo
  {pages} {27} (\bibinfo {year} {1975})}\BibitemShut {NoStop}%
\bibitem [{\citenamefont {Xi}\ \emph {et~al.}(2016)\citenamefont {Xi},
  \citenamefont {Wang}, \citenamefont {Zhao}, \citenamefont {Park},
  \citenamefont {Law}, \citenamefont {Berger}, \citenamefont {Forr{\'o}},
  \citenamefont {Shan},\ and\ \citenamefont {Mak}}]{xi16}%
  \BibitemOpen
  \bibfield  {author} {\bibinfo {author} {\bibfnamefont {X.}~\bibnamefont
  {Xi}}, \bibinfo {author} {\bibfnamefont {Z.}~\bibnamefont {Wang}}, \bibinfo
  {author} {\bibfnamefont {W.}~\bibnamefont {Zhao}}, \bibinfo {author}
  {\bibfnamefont {J.-H.}\ \bibnamefont {Park}}, \bibinfo {author}
  {\bibfnamefont {K.~T.}\ \bibnamefont {Law}}, \bibinfo {author} {\bibfnamefont
  {H.}~\bibnamefont {Berger}}, \bibinfo {author} {\bibfnamefont
  {L.}~\bibnamefont {Forr{\'o}}}, \bibinfo {author} {\bibfnamefont
  {J.}~\bibnamefont {Shan}}, \ and\ \bibinfo {author} {\bibfnamefont {K.~F.}\
  \bibnamefont {Mak}},\ }\bibfield  {title} {\bibinfo {title} {Ising pairing in
  superconducting NbSe$_{2}$ atomic layers},\ }\href {\doibase
  10.1038/nphys3538} {\bibfield  {journal} {\bibinfo  {journal} {Nature
  Physics}\ }\textbf {\bibinfo {volume} {12}},\ \bibinfo {pages} {139}
  (\bibinfo {year} {2016})}\BibitemShut {NoStop}%
\bibitem [{\citenamefont {Li}\ \emph {et~al.}()\citenamefont {Li},
  \citenamefont {Shen}, \citenamefont {Liu}, \citenamefont {Sha}, \citenamefont
  {Wang}, \citenamefont {Liu}, \citenamefont {Duan}, \citenamefont {Watanabe},
  \citenamefont {Taniguchi}, \citenamefont {Chen}, \citenamefont {Wang},
  \citenamefont {Zhong}, \citenamefont {Qian}, \citenamefont {Jiang},
  \citenamefont {Li}, \citenamefont {Yuan},\ and\ \citenamefont {Chen}}]{li26}%
  \BibitemOpen
  \bibfield  {author} {\bibinfo {author} {\bibfnamefont {Z.}~\bibnamefont
  {Li}}, \bibinfo {author} {\bibfnamefont {X.}~\bibnamefont {Shen}}, \bibinfo
  {author} {\bibfnamefont {K.}~\bibnamefont {Liu}}, \bibinfo {author}
  {\bibfnamefont {Y.}~\bibnamefont {Sha}}, \bibinfo {author} {\bibfnamefont
  {T.}~\bibnamefont {Wang}}, \bibinfo {author} {\bibfnamefont {F.}~\bibnamefont
  {Liu}}, \bibinfo {author} {\bibfnamefont {Q.}~\bibnamefont {Duan}}, \bibinfo
  {author} {\bibfnamefont {K.}~\bibnamefont {Watanabe}}, \bibinfo {author}
  {\bibfnamefont {T.}~\bibnamefont {Taniguchi}}, \bibinfo {author}
  {\bibfnamefont {P.}~\bibnamefont {Chen}}, \bibinfo {author} {\bibfnamefont
  {S.}~\bibnamefont {Wang}}, \bibinfo {author} {\bibfnamefont {R.}~\bibnamefont
  {Zhong}}, \bibinfo {author} {\bibfnamefont {D.}~\bibnamefont {Qian}},
  \bibinfo {author} {\bibfnamefont {S.}~\bibnamefont {Jiang}}, \bibinfo
  {author} {\bibfnamefont {Y.}~\bibnamefont {Li}}, \bibinfo {author}
  {\bibfnamefont {N.~F.~Q.}\ \bibnamefont {Yuan}}, \ and\ \bibinfo {author}
  {\bibfnamefont {G.}~\bibnamefont {Chen}},\ }\bibfield  {title} {\bibinfo
  {title} {Superconductivity in non-centrosymmetric rhombohedral NbSe$_{2}$},\
  }\href {https://arxiv.org/abs/2601.16475} {\bibinfo  {journal}
  {arXiv:2601.16475}\ }\BibitemShut {NoStop}%
\bibitem [{\citenamefont {Patra}\ \emph {et~al.}(2025)\citenamefont {Patra},
  \citenamefont {Agarwal}, \citenamefont {Verma}, \citenamefont {Manna},
  \citenamefont {Srivastava}, \citenamefont {Singh}, \citenamefont {Scheurer},
  \citenamefont {Singh},\ and\ \citenamefont {Singh}}]{patra25}%
  \BibitemOpen
\bibfield  {journal} {  }\bibfield  {author} {\bibinfo {author} {\bibfnamefont
  {C.}~\bibnamefont {Patra}}, \bibinfo {author} {\bibfnamefont
  {T.}~\bibnamefont {Agarwal}}, \bibinfo {author} {\bibfnamefont
  {R.}~\bibnamefont {Verma}}, \bibinfo {author} {\bibfnamefont
  {P.}~\bibnamefont {Manna}}, \bibinfo {author} {\bibfnamefont
  {S.}~\bibnamefont {Srivastava}}, \bibinfo {author} {\bibfnamefont {R.~S.}\
  \bibnamefont {Singh}}, \bibinfo {author} {\bibfnamefont {M.~S.}\ \bibnamefont
  {Scheurer}}, \bibinfo {author} {\bibfnamefont {B.}~\bibnamefont {Singh}}, \
  and\ \bibinfo {author} {\bibfnamefont {R.~P.}\ \bibnamefont {Singh}},\
  }\bibfield  {title} {\bibinfo {title} {Ising Superconductivity in Bulk
  Layered Noncentrosymmetric $4H\text{\ensuremath{-}}{\text{NbSe}}_{2}$},\
  }\href {\doibase 10.1103/m9xx-gk46} {\bibfield  {journal} {\bibinfo
  {journal} {Phys. Rev. Lett.}\ }\textbf {\bibinfo {volume} {135}},\ \bibinfo
  {pages} {216002} (\bibinfo {year} {2025})}\BibitemShut {NoStop}%
\bibitem [{\citenamefont {Martino}\ \emph {et~al.}(2026)\citenamefont
  {Martino}, \citenamefont {Arakcheeva}, \citenamefont {Berger}, \citenamefont
  {Prots}, \citenamefont {K{\"o}nig}, \citenamefont {Forr{\'o}},\ and\
  \citenamefont {Semeniuk}}]{marti26}%
  \BibitemOpen
  \bibfield  {author} {\bibinfo {author} {\bibfnamefont {E.}~\bibnamefont
  {Martino}}, \bibinfo {author} {\bibfnamefont {A.}~\bibnamefont {Arakcheeva}},
  \bibinfo {author} {\bibfnamefont {H.}~\bibnamefont {Berger}}, \bibinfo
  {author} {\bibfnamefont {Y.}~\bibnamefont {Prots}}, \bibinfo {author}
  {\bibfnamefont {M.}~\bibnamefont {K{\"o}nig}}, \bibinfo {author}
  {\bibfnamefont {L.}~\bibnamefont {Forr{\'o}}}, \ and\ \bibinfo {author}
  {\bibfnamefont {K.}~\bibnamefont {Semeniuk}},\ }\bibfield  {title} {\bibinfo
  {title} {Towards atomically-thin regime in bulk 4H-NbSe$_{2}$ with interlayer
  disorder},\ }\href {\doibase 10.1038/s41699-025-00659-w} {\bibfield
  {journal} {\bibinfo  {journal} {npj 2D Materials and Applications}\ }\textbf
  {\bibinfo {volume} {10}},\ \bibinfo {pages} {23} (\bibinfo {year}
  {2026})}\BibitemShut {NoStop}%
\bibitem [{\citenamefont {Nakata}\ \emph {et~al.}(2016)\citenamefont {Nakata},
  \citenamefont {Sugawara}, \citenamefont {Shimizu}, \citenamefont {Okada},
  \citenamefont {Han}, \citenamefont {Hitosugi}, \citenamefont {Ueno},
  \citenamefont {Sato},\ and\ \citenamefont {Takahashi}}]{nakat16}%
  \BibitemOpen
  \bibfield  {author} {\bibinfo {author} {\bibfnamefont {Y.}~\bibnamefont
  {Nakata}}, \bibinfo {author} {\bibfnamefont {K.}~\bibnamefont {Sugawara}},
  \bibinfo {author} {\bibfnamefont {R.}~\bibnamefont {Shimizu}}, \bibinfo
  {author} {\bibfnamefont {Y.}~\bibnamefont {Okada}}, \bibinfo {author}
  {\bibfnamefont {P.}~\bibnamefont {Han}}, \bibinfo {author} {\bibfnamefont
  {T.}~\bibnamefont {Hitosugi}}, \bibinfo {author} {\bibfnamefont
  {K.}~\bibnamefont {Ueno}}, \bibinfo {author} {\bibfnamefont {T.}~\bibnamefont
  {Sato}}, \ and\ \bibinfo {author} {\bibfnamefont {T.}~\bibnamefont
  {Takahashi}},\ }\bibfield  {title} {\bibinfo {title} {Monolayer 1T-NbSe$_{2}$
  as a Mott insulator},\ }\href {\doibase 10.1038/am.2016.157} {\bibfield
  {journal} {\bibinfo  {journal} {NPG Asia Materials}\ }\textbf {\bibinfo
  {volume} {8}},\ \bibinfo {pages} {e321} (\bibinfo {year} {2016})}\BibitemShut
  {NoStop}%
\bibitem [{\citenamefont {Nakata}\ \emph {et~al.}(2021)\citenamefont {Nakata},
  \citenamefont {Sugawara}, \citenamefont {Chainani}, \citenamefont {Oka},
  \citenamefont {Bao}, \citenamefont {Zhou}, \citenamefont {Chuang},
  \citenamefont {Cheng}, \citenamefont {Kawakami}, \citenamefont {Saruta},
  \citenamefont {Fukumura}, \citenamefont {Zhou}, \citenamefont {Takahashi},\
  and\ \citenamefont {Sato}}]{nakat21}%
  \BibitemOpen
  \bibfield  {author} {\bibinfo {author} {\bibfnamefont {Y.}~\bibnamefont
  {Nakata}}, \bibinfo {author} {\bibfnamefont {K.}~\bibnamefont {Sugawara}},
  \bibinfo {author} {\bibfnamefont {A.}~\bibnamefont {Chainani}}, \bibinfo
  {author} {\bibfnamefont {H.}~\bibnamefont {Oka}}, \bibinfo {author}
  {\bibfnamefont {C.}~\bibnamefont {Bao}}, \bibinfo {author} {\bibfnamefont
  {S.}~\bibnamefont {Zhou}}, \bibinfo {author} {\bibfnamefont {P.-Y.}\
  \bibnamefont {Chuang}}, \bibinfo {author} {\bibfnamefont {C.-M.}\
  \bibnamefont {Cheng}}, \bibinfo {author} {\bibfnamefont {T.}~\bibnamefont
  {Kawakami}}, \bibinfo {author} {\bibfnamefont {Y.}~\bibnamefont {Saruta}},
  \bibinfo {author} {\bibfnamefont {T.}~\bibnamefont {Fukumura}}, \bibinfo
  {author} {\bibfnamefont {S.}~\bibnamefont {Zhou}}, \bibinfo {author}
  {\bibfnamefont {T.}~\bibnamefont {Takahashi}}, \ and\ \bibinfo {author}
  {\bibfnamefont {T.}~\bibnamefont {Sato}},\ }\bibfield  {title} {\bibinfo
  {title} {Robust charge-density wave strengthened by electron correlations in
  monolayer 1T-TaSe$_{2}$ and 1T-NbSe$_{2}$},\ }\href {\doibase
  10.1038/s41467-021-26105-1} {\bibfield  {journal} {\bibinfo  {journal}
  {Nature Communications}\ }\textbf {\bibinfo {volume} {12}},\ \bibinfo {pages}
  {5873} (\bibinfo {year} {2021})}\BibitemShut {NoStop}%
\bibitem [{\citenamefont {Liu}\ \emph {et~al.}(2021{\natexlab{a}})\citenamefont
  {Liu}, \citenamefont {Qiao}, \citenamefont {Huang}, \citenamefont {Tang},
  \citenamefont {Ling}, \citenamefont {Zhang}, \citenamefont {Xia},
  \citenamefont {Liao}, \citenamefont {Shi}, \citenamefont {Mao}, \citenamefont
  {Zhu}, \citenamefont {L{\"u}},\ and\ \citenamefont {Fu}}]{liu21d}%
  \BibitemOpen
  \bibfield  {author} {\bibinfo {author} {\bibfnamefont {Z.-Y.}\ \bibnamefont
  {Liu}}, \bibinfo {author} {\bibfnamefont {S.}~\bibnamefont {Qiao}}, \bibinfo
  {author} {\bibfnamefont {B.}~\bibnamefont {Huang}}, \bibinfo {author}
  {\bibfnamefont {Q.-Y.}\ \bibnamefont {Tang}}, \bibinfo {author}
  {\bibfnamefont {Z.-H.}\ \bibnamefont {Ling}}, \bibinfo {author}
  {\bibfnamefont {W.-H.}\ \bibnamefont {Zhang}}, \bibinfo {author}
  {\bibfnamefont {H.-N.}\ \bibnamefont {Xia}}, \bibinfo {author} {\bibfnamefont
  {X.}~\bibnamefont {Liao}}, \bibinfo {author} {\bibfnamefont {H.}~\bibnamefont
  {Shi}}, \bibinfo {author} {\bibfnamefont {W.-H.}\ \bibnamefont {Mao}},
  \bibinfo {author} {\bibfnamefont {G.-L.}\ \bibnamefont {Zhu}}, \bibinfo
  {author} {\bibfnamefont {J.-T.}\ \bibnamefont {L{\"u}}}, \ and\ \bibinfo
  {author} {\bibfnamefont {Y.-S.}\ \bibnamefont {Fu}},\ }\bibfield  {title}
  {\bibinfo {title} {Charge Transfer Gap Tuning via Structural Distortion in
  Monolayer 1T-NbSe2},\ }\href {\doibase 10.1021/acs.nanolett.1c02348}
  {\bibfield  {journal} {\bibinfo  {journal} {Nano Letters}\ }\textbf {\bibinfo
  {volume} {21}},\ \bibinfo {pages} {7005} (\bibinfo {year}
  {2021}{\natexlab{a}})}\BibitemShut {NoStop}%
\bibitem [{\citenamefont {Liu}\ \emph {et~al.}(2021{\natexlab{b}})\citenamefont
  {Liu}, \citenamefont {Leveillee}, \citenamefont {Lu}, \citenamefont {Yu},
  \citenamefont {Kim}, \citenamefont {Tian}, \citenamefont {Shi}, \citenamefont
  {Lai}, \citenamefont {Zhang}, \citenamefont {Giustino},\ and\ \citenamefont
  {Shih}}]{liu21c}%
  \BibitemOpen
  \bibfield  {author} {\bibinfo {author} {\bibfnamefont {M.}~\bibnamefont
  {Liu}}, \bibinfo {author} {\bibfnamefont {J.}~\bibnamefont {Leveillee}},
  \bibinfo {author} {\bibfnamefont {S.}~\bibnamefont {Lu}}, \bibinfo {author}
  {\bibfnamefont {J.}~\bibnamefont {Yu}}, \bibinfo {author} {\bibfnamefont
  {H.}~\bibnamefont {Kim}}, \bibinfo {author} {\bibfnamefont {C.}~\bibnamefont
  {Tian}}, \bibinfo {author} {\bibfnamefont {Y.}~\bibnamefont {Shi}}, \bibinfo
  {author} {\bibfnamefont {K.}~\bibnamefont {Lai}}, \bibinfo {author}
  {\bibfnamefont {C.}~\bibnamefont {Zhang}}, \bibinfo {author} {\bibfnamefont
  {F.}~\bibnamefont {Giustino}}, \ and\ \bibinfo {author} {\bibfnamefont
  {C.-K.}\ \bibnamefont {Shih}},\ }\bibfield  {title} {\bibinfo {title}
  {Monolayer 1T-NbSe$_{2}$ as a 2D-correlated magnetic insulator},\ }\href
  {\doibase 10.1126/sciadv.abi6339} {\bibfield  {journal} {\bibinfo  {journal}
  {Science Advances}\ }\textbf {\bibinfo {volume} {7}},\ \bibinfo {pages}
  {eabi6339} (\bibinfo {year} {2021}{\natexlab{b}})}\BibitemShut {NoStop}%
\bibitem [{\citenamefont {Zhang}\ \emph {et~al.}(2024)\citenamefont {Zhang},
  \citenamefont {He}, \citenamefont {Zhang}, \citenamefont {Chen},
  \citenamefont {Jia}, \citenamefont {Hou}, \citenamefont {Ji}, \citenamefont
  {Yang}, \citenamefont {Zhang}, \citenamefont {Liu}, \citenamefont {Gao},
  \citenamefont {Jung},\ and\ \citenamefont {Wang}}]{zhang24b}%
  \BibitemOpen
  \bibfield  {author} {\bibinfo {author} {\bibfnamefont {Q.}~\bibnamefont
  {Zhang}}, \bibinfo {author} {\bibfnamefont {W.-Y.}\ \bibnamefont {He}},
  \bibinfo {author} {\bibfnamefont {Y.}~\bibnamefont {Zhang}}, \bibinfo
  {author} {\bibfnamefont {Y.}~\bibnamefont {Chen}}, \bibinfo {author}
  {\bibfnamefont {L.}~\bibnamefont {Jia}}, \bibinfo {author} {\bibfnamefont
  {Y.}~\bibnamefont {Hou}}, \bibinfo {author} {\bibfnamefont {H.}~\bibnamefont
  {Ji}}, \bibinfo {author} {\bibfnamefont {H.}~\bibnamefont {Yang}}, \bibinfo
  {author} {\bibfnamefont {T.}~\bibnamefont {Zhang}}, \bibinfo {author}
  {\bibfnamefont {L.}~\bibnamefont {Liu}}, \bibinfo {author} {\bibfnamefont
  {H.-J.}\ \bibnamefont {Gao}}, \bibinfo {author} {\bibfnamefont {T.~A.}\
  \bibnamefont {Jung}}, \ and\ \bibinfo {author} {\bibfnamefont
  {Y.}~\bibnamefont {Wang}},\ }\bibfield  {title} {\bibinfo {title} {Quantum
  spin liquid signatures in monolayer 1T-NbSe$_2$},\ }\href {\doibase
  10.1038/s41467-024-46612-1} {\bibfield  {journal} {\bibinfo  {journal}
  {Nature Communications}\ }\textbf {\bibinfo {volume} {15}},\ \bibinfo {pages}
  {2336} (\bibinfo {year} {2024})}\BibitemShut {NoStop}%
\bibitem [{\citenamefont {Shi}\ \emph {et~al.}(2024)\citenamefont {Shi},
  \citenamefont {Lin}, \citenamefont {Xiao}, \citenamefont {Weng},
  \citenamefont {Zhou}, \citenamefont {Long}, \citenamefont {Ding},
  \citenamefont {Liang}, \citenamefont {Huang}, \citenamefont {Chen},
  \citenamefont {Li},\ and\ \citenamefont {Zhang}}]{shi24}%
  \BibitemOpen
  \bibfield  {author} {\bibinfo {author} {\bibfnamefont {X.}~\bibnamefont
  {Shi}}, \bibinfo {author} {\bibfnamefont {D.}~\bibnamefont {Lin}}, \bibinfo
  {author} {\bibfnamefont {Z.}~\bibnamefont {Xiao}}, \bibinfo {author}
  {\bibfnamefont {Y.}~\bibnamefont {Weng}}, \bibinfo {author} {\bibfnamefont
  {H.}~\bibnamefont {Zhou}}, \bibinfo {author} {\bibfnamefont {X.}~\bibnamefont
  {Long}}, \bibinfo {author} {\bibfnamefont {Z.}~\bibnamefont {Ding}}, \bibinfo
  {author} {\bibfnamefont {F.}~\bibnamefont {Liang}}, \bibinfo {author}
  {\bibfnamefont {Y.}~\bibnamefont {Huang}}, \bibinfo {author} {\bibfnamefont
  {G.}~\bibnamefont {Chen}}, \bibinfo {author} {\bibfnamefont {K.}~\bibnamefont
  {Li}}, \ and\ \bibinfo {author} {\bibfnamefont {T.-Y.}\ \bibnamefont
  {Zhang}},\ }\bibfield  {title} {\bibinfo {title} {Exfoliation of bulk
  2H-MoS$_{2}$ into bilayer 1T-phase nanosheets via ether-induced
  superlattices},\ }\href {\doibase 10.1007/s12274-024-6446-3} {\bibfield
  {journal} {\bibinfo  {journal} {Nano Research}\ }\textbf {\bibinfo {volume}
  {17}},\ \bibinfo {pages} {5705} (\bibinfo {year} {2024})}\BibitemShut
  {NoStop}%
\bibitem [{\citenamefont {Kresse}\ and\ \citenamefont
  {Hafner}(1993)}]{kress93}%
  \BibitemOpen
  \bibfield  {author} {\bibinfo {author} {\bibfnamefont {G.}~\bibnamefont
  {Kresse}}\ and\ \bibinfo {author} {\bibfnamefont {J.}~\bibnamefont
  {Hafner}},\ }\bibfield  {title} {\bibinfo {title} {Ab initio molecular
  dynamics for liquid metals},\ }\href {\doibase 10.1103/PhysRevB.47.558}
  {\bibfield  {journal} {\bibinfo  {journal} {Phys. Rev. B}\ }\textbf {\bibinfo
  {volume} {47}},\ \bibinfo {pages} {558(R)} (\bibinfo {year}
  {1993})}\BibitemShut {NoStop}%
\bibitem [{\citenamefont {Kresse}\ and\ \citenamefont
  {Furthm\"uller}(1996)}]{kress96}%
  \BibitemOpen
  \bibfield  {author} {\bibinfo {author} {\bibfnamefont {G.}~\bibnamefont
  {Kresse}}\ and\ \bibinfo {author} {\bibfnamefont {J.}~\bibnamefont
  {Furthm\"uller}},\ }\bibfield  {title} {\bibinfo {title} {Efficient iterative
  schemes for ab initio total-energy calculations using a plane-wave basis
  set},\ }\href {\doibase 10.1103/PhysRevB.54.11169} {\bibfield  {journal}
  {\bibinfo  {journal} {Phys. Rev. B}\ }\textbf {\bibinfo {volume} {54}},\
  \bibinfo {pages} {11169} (\bibinfo {year} {1996})}\BibitemShut {NoStop}%
\bibitem [{\citenamefont {Bl\"ochl}(1994)}]{bloch94}%
  \BibitemOpen
  \bibfield  {author} {\bibinfo {author} {\bibfnamefont {P.~E.}\ \bibnamefont
  {Bl\"ochl}},\ }\bibfield  {title} {\bibinfo {title} {Projector augmented-wave
  method},\ }\href {\doibase 10.1103/PhysRevB.50.17953} {\bibfield  {journal}
  {\bibinfo  {journal} {Phys. Rev. B}\ }\textbf {\bibinfo {volume} {50}},\
  \bibinfo {pages} {17953} (\bibinfo {year} {1994})}\BibitemShut {NoStop}%
\bibitem [{\citenamefont {Kresse}\ and\ \citenamefont
  {Joubert}(1999)}]{kress99}%
  \BibitemOpen
  \bibfield  {author} {\bibinfo {author} {\bibfnamefont {G.}~\bibnamefont
  {Kresse}}\ and\ \bibinfo {author} {\bibfnamefont {D.}~\bibnamefont
  {Joubert}},\ }\bibfield  {title} {\bibinfo {title} {From ultrasoft
  pseudopotentials to the projector augmented-wave method},\ }\href {\doibase
  10.1103/PhysRevB.59.1758} {\bibfield  {journal} {\bibinfo  {journal} {Phys.
  Rev. B}\ }\textbf {\bibinfo {volume} {59}},\ \bibinfo {pages} {1758}
  (\bibinfo {year} {1999})}\BibitemShut {NoStop}%
\bibitem [{\citenamefont {Perdew}\ \emph {et~al.}(1996)\citenamefont {Perdew},
  \citenamefont {Burke},\ and\ \citenamefont {Ernzerhof}}]{perde96}%
  \BibitemOpen
  \bibfield  {author} {\bibinfo {author} {\bibfnamefont {J.~P.}\ \bibnamefont
  {Perdew}}, \bibinfo {author} {\bibfnamefont {K.}~\bibnamefont {Burke}}, \
  and\ \bibinfo {author} {\bibfnamefont {M.}~\bibnamefont {Ernzerhof}},\
  }\bibfield  {title} {\bibinfo {title} {Generalized Gradient Approximation
  Made Simple},\ }\href {\doibase 10.1103/PhysRevLett.77.3865} {\bibfield
  {journal} {\bibinfo  {journal} {Phys. Rev. Lett.}\ }\textbf {\bibinfo
  {volume} {77}},\ \bibinfo {pages} {3865} (\bibinfo {year}
  {1996})}\BibitemShut {NoStop}%
\bibitem [{\citenamefont {Perdew}\ \emph {et~al.}(1997)\citenamefont {Perdew},
  \citenamefont {Burke},\ and\ \citenamefont {Ernzerhof}}]{perde97}%
  \BibitemOpen
  \bibfield  {author} {\bibinfo {author} {\bibfnamefont {J.~P.}\ \bibnamefont
  {Perdew}}, \bibinfo {author} {\bibfnamefont {K.}~\bibnamefont {Burke}}, \
  and\ \bibinfo {author} {\bibfnamefont {M.}~\bibnamefont {Ernzerhof}},\
  }\bibfield  {title} {\bibinfo {title} {Generalized Gradient Approximation
  Made Simple [Phys. Rev. Lett. 77, 3865 (1996)]},\ }\href {\doibase
  10.1103/PhysRevLett.78.1396} {\bibfield  {journal} {\bibinfo  {journal}
  {Phys. Rev. Lett.}\ }\textbf {\bibinfo {volume} {78}},\ \bibinfo {pages}
  {1396} (\bibinfo {year} {1997})}\BibitemShut {NoStop}%
\bibitem [{\citenamefont {Liu}\ \emph {et~al.}(2024)\citenamefont {Liu},
  \citenamefont {Jin}, \citenamefont {Zhang}, \citenamefont {Fan},
  \citenamefont {Guo}, \citenamefont {Qin}, \citenamefont {Zhu}, \citenamefont
  {Yang}, \citenamefont {Zhang}, \citenamefont {Huang},\ and\ \citenamefont
  {Fu}}]{liu24c}%
  \BibitemOpen
  \bibfield  {author} {\bibinfo {author} {\bibfnamefont {Z.-Y.}\ \bibnamefont
  {Liu}}, \bibinfo {author} {\bibfnamefont {H.}~\bibnamefont {Jin}}, \bibinfo
  {author} {\bibfnamefont {Y.}~\bibnamefont {Zhang}}, \bibinfo {author}
  {\bibfnamefont {K.}~\bibnamefont {Fan}}, \bibinfo {author} {\bibfnamefont
  {T.-F.}\ \bibnamefont {Guo}}, \bibinfo {author} {\bibfnamefont {H.-J.}\
  \bibnamefont {Qin}}, \bibinfo {author} {\bibfnamefont {L.-F.}\ \bibnamefont
  {Zhu}}, \bibinfo {author} {\bibfnamefont {L.-Z.}\ \bibnamefont {Yang}},
  \bibinfo {author} {\bibfnamefont {W.-H.}\ \bibnamefont {Zhang}}, \bibinfo
  {author} {\bibfnamefont {B.}~\bibnamefont {Huang}}, \ and\ \bibinfo {author}
  {\bibfnamefont {Y.-S.}\ \bibnamefont {Fu}},\ }\bibfield  {title} {\bibinfo
  {title} {Charge-density wave mediated quasi-one-dimensional Kondo lattice in
  stripe-phase monolayer 1T-NbSe$_2$},\ }\href {\doibase
  10.1038/s41467-024-45335-7} {\bibfield  {journal} {\bibinfo  {journal}
  {Nature Communications}\ }\textbf {\bibinfo {volume} {15}},\ \bibinfo {pages}
  {1039} (\bibinfo {year} {2024})}\BibitemShut {NoStop}%
\bibitem [{\citenamefont {Grimme}\ \emph {et~al.}(2010)\citenamefont {Grimme},
  \citenamefont {Antony}, \citenamefont {Ehrlich},\ and\ \citenamefont
  {Krieg}}]{grimm10}%
  \BibitemOpen
  \bibfield  {author} {\bibinfo {author} {\bibfnamefont {S.}~\bibnamefont
  {Grimme}}, \bibinfo {author} {\bibfnamefont {J.}~\bibnamefont {Antony}},
  \bibinfo {author} {\bibfnamefont {S.}~\bibnamefont {Ehrlich}}, \ and\
  \bibinfo {author} {\bibfnamefont {H.}~\bibnamefont {Krieg}},\ }\bibfield
  {title} {\bibinfo {title} {A consistent and accurate ab initio
  parametrization of density functional dispersion correction (DFT-D) for the
  94 elements H-Pu},\ }\href {\doibase 10.1063/1.3382344} {\bibfield  {journal}
  {\bibinfo  {journal} {The Journal of Chemical Physics}\ }\textbf {\bibinfo
  {volume} {132}},\ \bibinfo {pages} {154104} (\bibinfo {year}
  {2010})}\BibitemShut {NoStop}%
\bibitem [{\citenamefont {Meyer}\ \emph {et~al.}(1975)\citenamefont {Meyer},
  \citenamefont {Howard}, \citenamefont {Stewart}, \citenamefont {Acrivos},\
  and\ \citenamefont {Geballe}}]{meyer75}%
  \BibitemOpen
  \bibfield  {author} {\bibinfo {author} {\bibfnamefont {S.~F.}\ \bibnamefont
  {Meyer}}, \bibinfo {author} {\bibfnamefont {R.~E.}\ \bibnamefont {Howard}},
  \bibinfo {author} {\bibfnamefont {G.~R.}\ \bibnamefont {Stewart}}, \bibinfo
  {author} {\bibfnamefont {J.~V.}\ \bibnamefont {Acrivos}}, \ and\ \bibinfo
  {author} {\bibfnamefont {T.~H.}\ \bibnamefont {Geballe}},\ }\bibfield
  {title} {\bibinfo {title} {Properties of intercalated 2H‐NbSe$_2$,
  4H$_b$‐TaS$_2$, and 1T‐TaS$_2$},\ }\href {\doibase 10.1063/1.430342}
  {\bibfield  {journal} {\bibinfo  {journal} {The Journal of Chemical Physics}\
  }\textbf {\bibinfo {volume} {62}},\ \bibinfo {pages} {4411} (\bibinfo {year}
  {1975})}\BibitemShut {NoStop}%
\bibitem [{\citenamefont {Naik}\ \emph {et~al.}(2019)\citenamefont {Naik},
  \citenamefont {Pradhan}, \citenamefont {Bhat}, \citenamefont {Behera},
  \citenamefont {Kumar}, \citenamefont {Samal},\ and\ \citenamefont
  {Samal}}]{naik19}%
  \BibitemOpen
  \bibfield  {author} {\bibinfo {author} {\bibfnamefont {S.}~\bibnamefont
  {Naik}}, \bibinfo {author} {\bibfnamefont {G.~K.}\ \bibnamefont {Pradhan}},
  \bibinfo {author} {\bibfnamefont {S.~G.}\ \bibnamefont {Bhat}}, \bibinfo
  {author} {\bibfnamefont {B.~C.}\ \bibnamefont {Behera}}, \bibinfo {author}
  {\bibfnamefont {P.~A.}\ \bibnamefont {Kumar}}, \bibinfo {author}
  {\bibfnamefont {S.~L.}\ \bibnamefont {Samal}}, \ and\ \bibinfo {author}
  {\bibfnamefont {D.}~\bibnamefont {Samal}},\ }\bibfield  {title} {\bibinfo
  {title} {The effect of Sn intercalation on the superconducting properties of
  2H-NbSe$_{2}$},\ }\href {\doibase
  https://doi.org/10.1016/j.physc.2019.02.011} {\bibfield  {journal} {\bibinfo
  {journal} {Physica C: Superconductivity and its Applications}\ }\textbf
  {\bibinfo {volume} {561}},\ \bibinfo {pages} {18 } (\bibinfo {year}
  {2019})}\BibitemShut {NoStop}%
\bibitem [{\citenamefont {Di~Salvo}\ \emph {et~al.}(1976)\citenamefont
  {Di~Salvo}, \citenamefont {Wilson},\ and\ \citenamefont {Waszczak}}]{disa76}%
  \BibitemOpen
  \bibfield  {author} {\bibinfo {author} {\bibfnamefont {F.~J.}\ \bibnamefont
  {Di~Salvo}}, \bibinfo {author} {\bibfnamefont {J.~A.}\ \bibnamefont
  {Wilson}}, \ and\ \bibinfo {author} {\bibfnamefont {J.~V.}\ \bibnamefont
  {Waszczak}},\ }\bibfield  {title} {\bibinfo {title} {Localization of
  Conduction Electrons by Fe, Co, and Ni in 1T-Ta${\mathrm{S}}_{2}$ and
  1T-Ta${\mathrm{Se}}_{2}$},\ }\href {\doibase 10.1103/PhysRevLett.36.885}
  {\bibfield  {journal} {\bibinfo  {journal} {Phys. Rev. Lett.}\ }\textbf
  {\bibinfo {volume} {36}},\ \bibinfo {pages} {885} (\bibinfo {year}
  {1976})}\BibitemShut {NoStop}%
\bibitem [{\citenamefont {Wilson}\ \emph {et~al.}(1975)\citenamefont {Wilson},
  \citenamefont {Salvo},\ and\ \citenamefont {Mahajan}}]{wilso75a}%
  \BibitemOpen
  \bibfield  {author} {\bibinfo {author} {\bibfnamefont {J.}~\bibnamefont
  {Wilson}}, \bibinfo {author} {\bibfnamefont {F.~D.}\ \bibnamefont {Salvo}}, \
  and\ \bibinfo {author} {\bibfnamefont {S.}~\bibnamefont {Mahajan}},\
  }\bibfield  {title} {\bibinfo {title} {Charge-density waves and superlattices
  in the metallic layered transition metal dichalcogenides},\ }\href {\doibase
  10.1080/00018737500101391} {\bibfield  {journal} {\bibinfo  {journal}
  {Advances in Physics}\ }\textbf {\bibinfo {volume} {24}},\ \bibinfo {pages}
  {117} (\bibinfo {year} {1975})}\BibitemShut {NoStop}%
\bibitem [{\citenamefont {Law}\ and\ \citenamefont {Lee}(2017)}]{law17}%
  \BibitemOpen
  \bibfield  {author} {\bibinfo {author} {\bibfnamefont {K.~T.}\ \bibnamefont
  {Law}}\ and\ \bibinfo {author} {\bibfnamefont {P.~A.}\ \bibnamefont {Lee}},\
  }\bibfield  {title} {\bibinfo {title} {1T-TaS$_{2}$ as a quantum spin
  liquid},\ }\href {\doibase 10.1073/pnas.1706769114} {\bibfield  {journal}
  {\bibinfo  {journal} {Proceedings of the National Academy of Sciences}\
  }\textbf {\bibinfo {volume} {114}},\ \bibinfo {pages} {6996} (\bibinfo {year}
  {2017})}\BibitemShut {NoStop}%
\bibitem [{\citenamefont {Zhang}\ and\ \citenamefont {Wu}(2023)}]{zhang23c}%
  \BibitemOpen
  \bibfield  {author} {\bibinfo {author} {\bibfnamefont {W.}~\bibnamefont
  {Zhang}}\ and\ \bibinfo {author} {\bibfnamefont {J.}~\bibnamefont {Wu}},\
  }\bibfield  {title} {\bibinfo {title} {Stacking order and driving forces in
  the layered charge density wave phase of 1T-MX$_2$ (M = Nb, Ta and X = S,
  Se)},\ }\href {\doibase 10.1088/2053-1591/acc997} {\bibfield  {journal}
  {\bibinfo  {journal} {Materials Research Express}\ }\textbf {\bibinfo
  {volume} {10}},\ \bibinfo {pages} {046302} (\bibinfo {year}
  {2023})}\BibitemShut {NoStop}%
\bibitem [{\citenamefont {Pereira}\ and\ \citenamefont
  {Liang}(1982)}]{perei82}%
  \BibitemOpen
  \bibfield  {author} {\bibinfo {author} {\bibfnamefont {C.~M.}\ \bibnamefont
  {Pereira}}\ and\ \bibinfo {author} {\bibfnamefont {W.~Y.}\ \bibnamefont
  {Liang}},\ }\bibfield  {title} {\bibinfo {title} {Raman studies of the normal
  phase of 2H-NbSe$_2$},\ }\href {\doibase 10.1088/0022-3719/15/27/009}
  {\bibfield  {journal} {\bibinfo  {journal} {Journal of Physics C: Solid State
  Physics}\ }\textbf {\bibinfo {volume} {15}},\ \bibinfo {pages} {L991}
  (\bibinfo {year} {1982})}\BibitemShut {NoStop}%
\end{thebibliography}

\end{document}